\documentclass[twocolumn,showpacs,preprintnumbers,amsmath,amssymb]{revtex4}

\usepackage{graphicx}
\usepackage{dcolumn}
\usepackage{bm}

\begin{document}

\title{Self-Organization Induced Scale-Free Networks}
\author{Gang Yan$^{1}$}
\email{russell0123@ustc.edu}
\author{Tao Zhou$^{1,2}$}
\email{zhutou@ustc.edu}
\author{Ying-Di Jin$^{2}$}
\author{Zhong-Qian Fu$^{1}$}
 \affiliation{
Department of Electronic Science and Technology,\\
University of Science and Technology of China,\\
Hefei Anhui, 230026, PR China\\
Department of Modern Physics, \\
University of Science and Technology of China, \\
Hefei, Anhui, 230026, PR China}

\date{\today}

\begin{abstract}

What is the underlying mechanism leading to power-law degree
distributions of many natural and artificial networks is still at
issue. We consider that scale-free networks emerges from
self-organizing process, and such a evolving model is introduced
in this letter. At each time step, a new node is added to the
network and connect to some existing nodes randomly, instead of
``preferential attachment" introduced by Barab\'{a}si and Albert,
and then the new node will connect with its neighbors' neighbors
at a fixed probability, which is natural to collaboration networks
and social networks of acquaintance or other relations between
individuals. The simulation results show that those networks
generated from our model are scale-free networks with
satisfactorily large clustering coefficient.
\end{abstract}


\maketitle

The last few years have witnessed a tremendous activity devoted to
the characterization and understanding of complex networks
\cite{Reviews1, Reviews2, Reviews3}, which arise in a vast number
of natural and artificial systems, such as Internet
\cite{Internet1,Internet2,Internet3}, the World-Wide Web
\cite{www1,www2}, social networks
\cite{social1,social2,social3,social4}, airports network
\cite{airline1,airline2}, food webs \cite{foodwebs1,foodwebs2},
biological interacting networks \cite{bio1,bio2,bio3} and so
forth. Particularly, much attention has been dedicated recently to
the study of scale-free networks, namely, networks that display a
power-law degree distributions, $P(k)\propto k^{-\gamma}$, where
$k$ is the connectivity (degree) \cite{links}. And various
proposals for dynamical evolution of scale-free networks has been
introduced. Roughly, these models can be classified into two main
scenarios \cite{classified}. One is under the mechanism
``preferential attachment", which means new vertices are
preferentially attached to existing vertices with large number of
neighbors \cite{BArelated1,BArelated2,BArelated3,BArelated4}, and
a related scenario is found in the protein duplication model
\cite{protein}. Another is a balance between a modelled tendency
to form hubs against an entropy pressure towards a random networks
\cite{BalanceModel1,BalanceModel2,BalanceModel3}.

\begin{figure}
\scalebox{0.45}[0.5]{\includegraphics{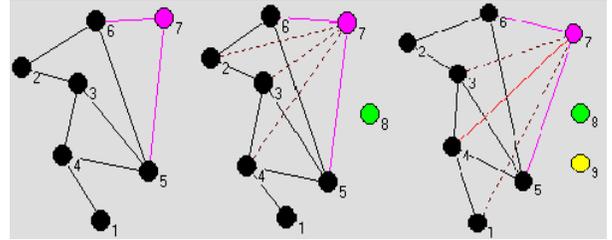}}
\caption{\label{fig:epsart} A schematic representation of the
evolving rules for the case $m=2,G=2$. At each time step, a new
node is added to the network. For each new node, it will be active
along $G+1$ steps. In the first time step(i.e. the step when it
was added to the network), it will randomly choose $m$ nodes to be
its neighbors. Then, in the following $G$ steps, with probability
$p$, this node will connect to the neighbors of the nodes which
connected to it at the last step. To make it easier to understand,
we draw the figure 1 as a sketch map for the case $m=2,G=2$.
Assume node \emph{7} connected to two nodes \emph{5 and 6}(left),
then it will link to the neighbors of \emph{5 and 6}, i.e.
\emph{2, 3, 4}, with probability $p$. Obviously, if some node is
common neighbor of \emph{5 and 6}, the probability is 2$p$. At the
same step a new node \emph{8} is added to the network by linking
to \emph{m=2} nodes randomly (omitted in Fig.1) (mid). Suppose
node \emph{7} connected to node \emph{4} at last step, then it
links to neighbors of \emph{4}, i.e. \emph{3, 1}, with the same
probability $p$ (right). The operation for each node lasts for $G$
steps similarly.}
\end{figure}

However, what is the underlying mechanism leading to power-law
degree distributions is still at issue. Differing from the two
class of mechanisms mentioned above, we consider that
self-organization maybe the fundamental mechanism which leads to
power-law distribution of degree, and a model including such a
mechanism is introduced in this letter. We think that the node
newly added to the network will connect to some existing nodes
randomly, not preferentially. For instance, considering the
scientist collaboration networks, most of the times, a scholar
comes into a new field not because he correlated with some famous
people in this field , but he read some experts' papers by chance
and then connected with them or some of his partners were in this
field already. Thus, a relatively new node having linked to some
other nodes in the network will connect with its neighbors'
neighbors, whose evidences can be found in collaboration and
friendship networks. Therefore, we propose an evolving model
following the rules:
\\

1. Starting with a small number ($m_{{\rm 0}}$) of nodes which are
global connected with each other. At each time step, a new node
$i$ is added to the networks, and connect to $m$ nodes randomly.
\\

2. The node $i$ will link with probability $p$ to the neighbors of
the nodes which connected to it at the last step. And at the next
$G-1$ steps ($G$ is a fixed integer), let node $i$ execute similar
operation (as \emph{Fig.1} shows).
\\

\begin{figure}
\scalebox{0.41}[0.5]{\includegraphics{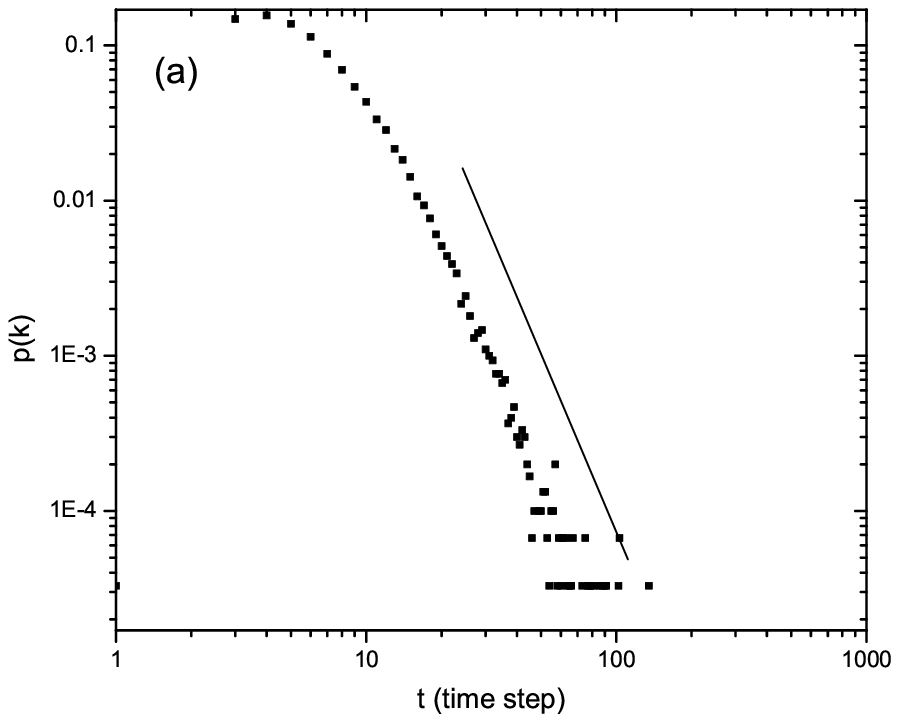}}
\scalebox{0.41}[0.5]{\includegraphics{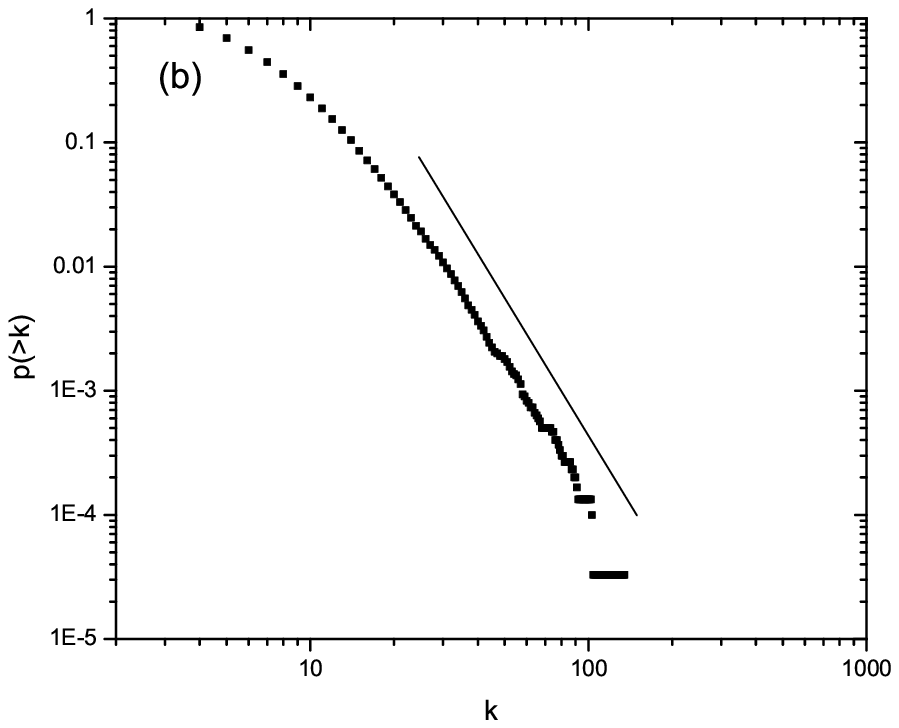}}
\scalebox{0.41}[0.5]{\includegraphics{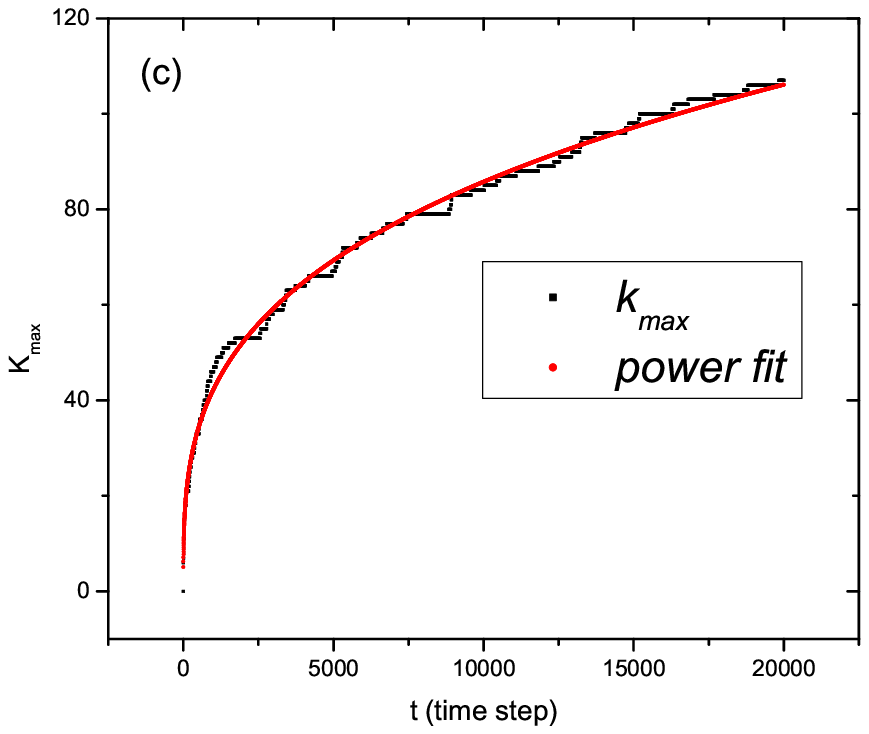}}
\scalebox{0.41}[0.54]{\includegraphics{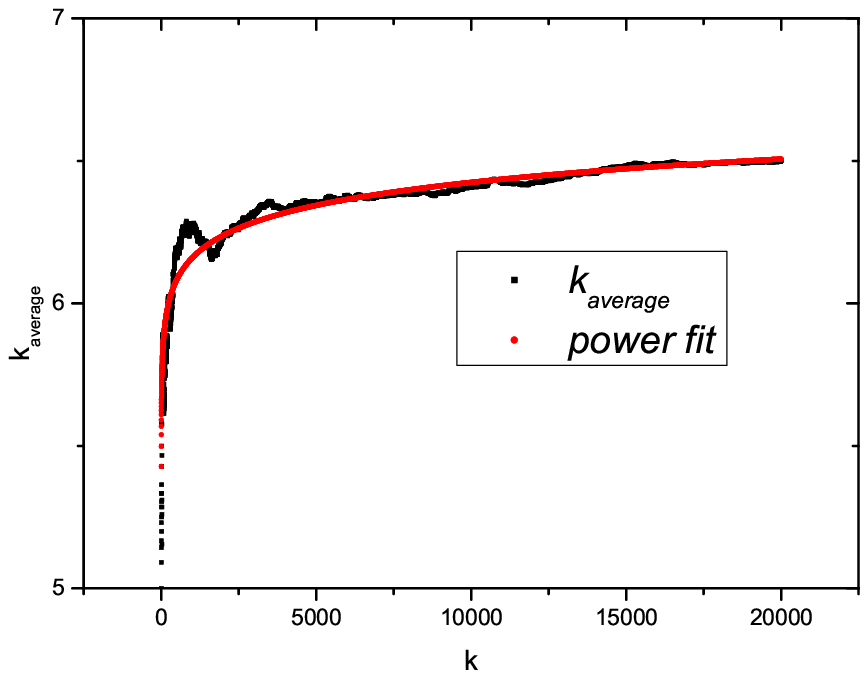}} \caption{The
statistical and evolutive characterization of degree, where $m=3,
p=0.06, G=2, N=2\times 10^4$: (a)degree distribution $p(k)$, (b)
normalized accumulative degree distribution
$P(k)=\int^\infty_{k}kp(k)dk/k$, (c)time evolution of maximum
degree $k_{{\rm max}}$ and (d)average degree $k_{{\rm average}}$.
Obviously, $p(k)\propto k^{-\gamma}$ , where $\gamma=2.85 \pm
0.07$. Power fitness ($a\ast t^{{\rm b}}$) of $k_{{\rm max}}$ and
$k_{{\rm average}}$ are represented, with $a=5.06 \pm 0.06,
b=0.307 \pm 0.005$ and $a=5.42 \pm 0.06, b=0.018 \pm 0.001$
respectively, i.e. $k_{{\rm max}} \propto N^{{\rm 0.307}}$,
$k_{{\rm average}} \propto N^{{\rm 0.018}}$. The number of edges
of the networks $E=\frac{N\ast k_{{\rm average}}}{2}\propto
N^{{\rm 1.018}}$, in other words, the networks are sparse.}
\end{figure}

\begin{figure}
\scalebox{0.8}[0.8]{\includegraphics{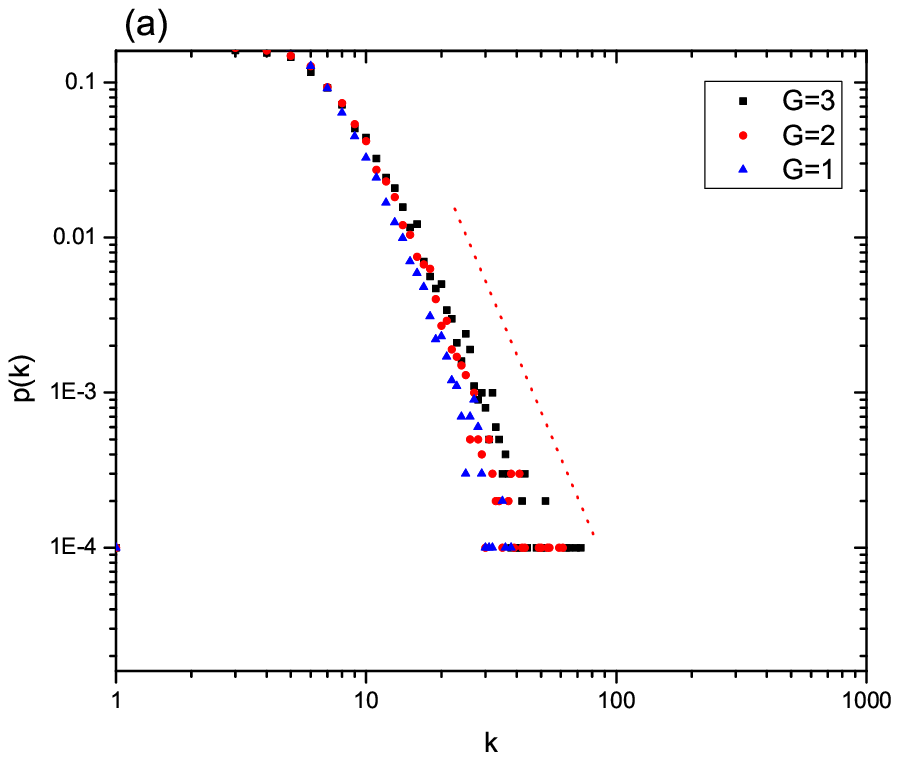}}
\scalebox{0.8}[0.8]{\includegraphics{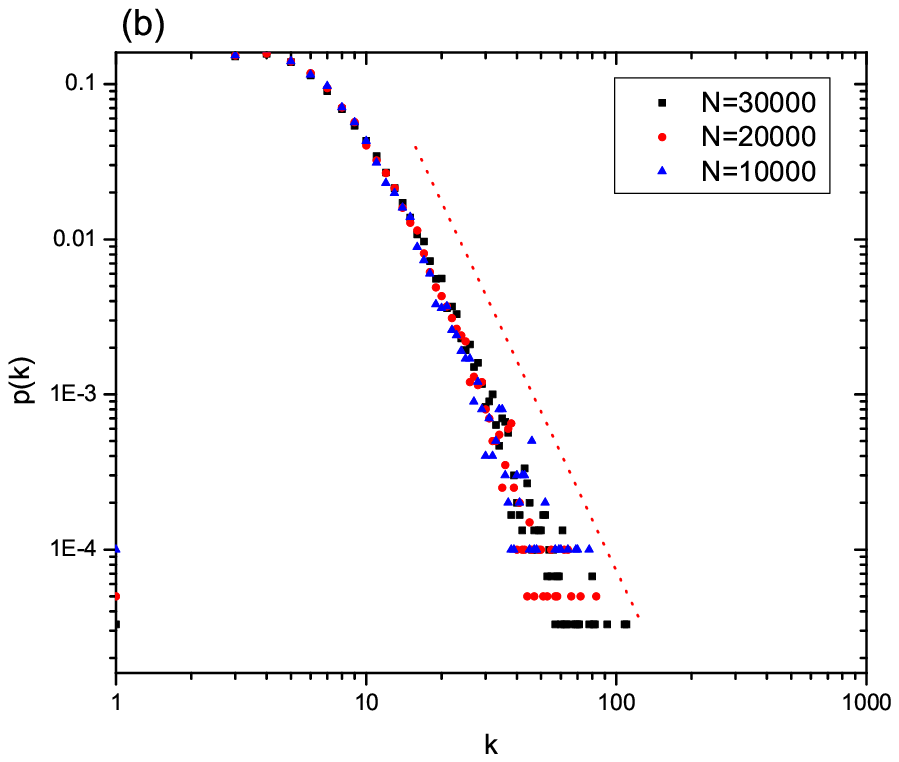}} \caption{Degree
distribution for different $G$ and $N$: (a)$G$=1, 2, 3, $N=10^4$;
(b) $N=10^4, 2\times10^4, 3\times10^4$, $G$=2, where $m$=2,
$p$=0.06. It shows that the value of $G$ and $N$ impacts little
the topology of network, i.e. different value of $G$ and $N$ leads
to almost the same power-law behavior of degree distribution.}
\end{figure}

For simplification, the parameters $m$, $p$ and $G$ are constant.
Obviously, rule\emph{2} is a self-organized process that impacts
the topological structure of the networks. \emph{Fig.2} shows a
typical scale-free network based on our model with $N=2\times10^4$
nodes. The exponent $\gamma$ is a little smaller than 3.0, which
is close to empirical study. And time evolution of maximum degree
$k_{{\rm max}}$ and average degree $k_{{\rm average}}$ are
exhibited in \emph{Fig.2c} and \emph{Fig.2d}, respectively. It can
be found that $k_{{\rm max}}\propto N^{{\rm 0.307}}$, $k_{{\rm
average}}\propto N^{{\rm 0.018}}$, thus the total number of edges
in the network is proportionable to $N^{{\rm 1+0.018}}$, in other
words, the networks is sparse. Moreover, there is evidence to
suggest that in real-world networks, e.g. World Wide Web, the
average degree of nodes is increasing with time \cite{Reviews3}.
Our model accords with that very well.

\begin{figure}
\scalebox{0.8}[0.8]{\includegraphics{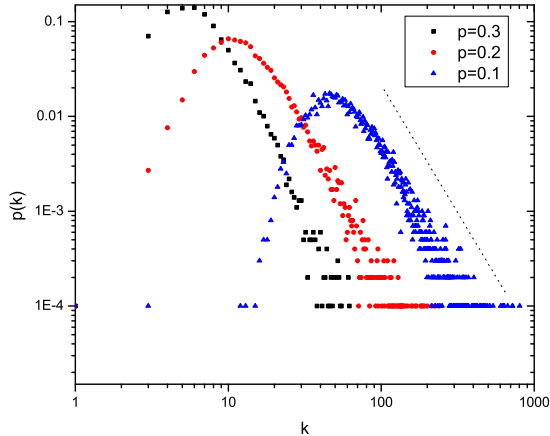}} \caption{The
degree distribution $p(k)$ for networks with $p$=0.1, 0.2, 0.3,
where $m$=2, $G$=1, $N=10^4$. It shows that for different
probability $p$, the degree distributions follows power-law
behaviors $p(k)\propto k^{-\gamma}$ with almost the same exponent
$\gamma=2.90 \pm 0.09$, and for larger $p$, the rang of power-law
distribution gets narrower.}
\end{figure}

\begin{figure}
\scalebox{0.8}[0.8]{\includegraphics{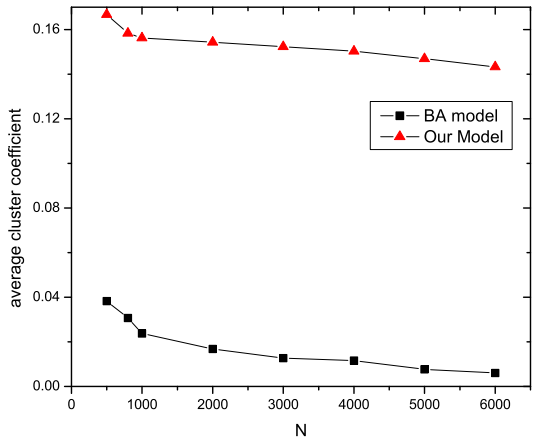}} \caption{
Clustering coefficient of networks with different size, $N$, based
on our model(triangle) and BA model(square). Obviously, the one of
our model is much larger.}
\end{figure}

In the following we will discuss how the parameters $G$ and $p$
impact the topology of networks. In fig.3a one can see that for
different $G$ the model leads to almost the same degree
distribution, i.e. the value of $G$ impacts little the topology of
networks. One can easily check that even for large $G$, which is
close to real-world instances that people usually have wide
communication with others, the networks based on our model are
still of scale-free properties. This extensive result is very
significant. However, when $G$ is too large (e.g. $G \geq 10$) the
computer is unable to work it out. Figure 3b exhibits that $p(k)$
is independent of time and subsequently independent of network
size $N=m_{{\rm 0}}+t$, which indicates that despite its
continuous growth, the network self-organizes to a scale-free
stationary state.

As illuminated above, for small values of $p$, e.g. 0.06, 0.08
\emph{et al}, the self-organization processes induce scale-free
networks. However, when $p$ is getting larger, that's to say, each
node newly added to the network connects to its neighbors'
neighbors with larger probability, the number of nodes with small
degree will become fewer (as shown in \emph{Fig.4}). The results
indicate that the rang of power-law degree distribution will get
narrower as $p$ increases. Let $p$=0, the network has exponential
degree distribution. The analysis is simple. We can draw the
master equation as follows:
\begin{equation}
N(k,t+1)=N(k,t)+\frac{N(k-1,t)}{t+m_{{\rm 0}}}\times
m-\frac{N(k,t)}{t+m_{{\rm 0}}}\times m
\end{equation}
where $N(k,t)$ is the number of nodes with degree $k$ at time step
$t$. Since $N(k,t)\sim p(k)t$ and $t\gg m_{{\rm 0}}$ for large
$t$, seeking solution of this form we can get that $p(k)\propto
e^{{\rm -\frac{k}{m}}}$. How $p$ impacts the degree distribution
detailedly is significant and will be discussed in the future
work.

It is well known one of the shortcomings of BA's model is that the
clustering coefficient is small and decreases with the increasing
of network size, following approximately a power law $C\sim
N^{{\rm -0.75}}$ \cite{Reviews1}. In \emph{Fig.5}, we report the
clustering coefficient of the networks based on our model, which
is much larger than that on BA's model. This is not difficult to
explain. Suppose node $h1$ and $h2$ are neighbors of a common node
$h3$, then according to rule\emph{2} mentioned above, node $h1$
will link to node $h2$ with probability $p$ if the former one is
added to the network earlier than the later one. This induces that
the edges among neighbors of node $h3$ will become more, thus the
clustering coefficient will get larger.

In summary, we propose a new model, based on ``randomly
attachment" and self-organization, which leads to a scale-free
degree distribution of networks. The mechanism is different from
preferential attachment mechanism where a scale-free distribution
are generated during gradual growth of hubs. To investigate the
``randomly attachment" mechanism is very important, since there
are some networks that appear to have power-law degree
distributions, but for which preferential attachment is clearly
not an appropriate model\cite{bio1,bio3,Fell2000,Sole2003}. The
simulation results exhibit that, rather than BA networks, the
networks generated from our model are of larger clustering, which
is close to the real-life networks. Furthermore, we suggest that
self-organizing process plays a major role in many real-life
networks such as collaboration networks, social networks of
acquaintance or other relations between individuals and so on. We
believe that many useful scale-free networks can be constructed
using our approach.

\end{document}